\documentclass[useAMS,usenatbib,usegraphicx,psfig]{mn2e}

\def\m@th{\mathsurround=0pt } 
\def\eqalign#1{\null\,\vcenter{\openup1\jot\m@th
 \ialign{\strut\hfil$\displaystyle{##}$&$\displaystyle{{}##}$\hfil
 \crcr#1\crcr}}\,}

\title[Colour gradients in E/S0 galaxies]
{Internal colour gradients for E/S0 galaxies in Abell 2218}
\author[Roberto De Propris et~al.]{
\parbox[t]{\textwidth}{
Roberto De Propris$^{1,2,3}$\thanks{R.DePropris@bristol.ac.uk},
Matthew Colless$^3$,
Simon P.\ Driver$^2$,
Michael B.\ Pracy$^4$,
Warrick J.\ Couch$^4$
}
\vspace*{6pt} \\
$^1$Astrophysics Group, Department of Physics, University of Bristol,
    Tyndall Avenue, Bristol, BS8 1TL, United Kingdom\\
$^2$Research School of Astronomy \& Astrophysics, The Australian 
    National University, Weston Creek, ACT 2611, Australia \\
$^3$Anglo-Australian Observatory, P.O.\ Box 296, Epping, NSW 2111,
    Australia\\  
$^4$Department of Astrophysics, University of New South Wales, Sydney, 
    NSW 2052, Australia \\
}

\begin{document}
\date{}
\pagerange{\pageref{firstpage}--\pageref{lastpage}}
\pubyear{2004}
\maketitle
\label{firstpage}

\begin{abstract}
  
  We determine colour gradients of $-0.15 \pm 0.08$ magnitudes per
  decade in radius in F450W$-$F606W and $-0.07 \pm 0.06$ magnitudes per
  decade in radius in F606W$-$F814W for a sample of 22 E/S0 galaxies in
  Abell 2218. These gradients are consistent with the existence of a
  mild ($\sim -0.3$ dex per decade in radius) gradient in metal abundance, 
  (cf. previous work at lower and higher redshift for field and cluster 
  galaxies). The size of the observed gradients is found to be independent 
  of luminosity over a range spanning $M^*-1$ to $M^*+1.5$ and also to be 
  independent of morphological type. These results suggest a fundamental 
  similarity in the distributions of stellar populations in ellipticals 
  and the bulges of lenticular galaxies. These results are not consistent 
  with simple models of either monolithic collapse or hierarchical mergers.

\end{abstract}


\section{Introduction}

In Hubble's (1936) classification scheme for galaxies, lenticular (S0)
galaxies first appear as a transitional class between the ellipticals
and the two branches (barred and unbarred) of the spirals, and, despite
much evidence to the contrary, it has often been assumed that the Hubble
`fork' describes an evolutionary scheme. For instance, this assumption
informs the popular idea that collisions of proto-disks in the early
universe lead to the formation of elliptical galaxies or that quenching
of star formation in field spirals as they fall into dense regions
leads, by a variety of mechanisms, to their transformation into S0
galaxies. This process, first proposed by Larson, Tinsley \& Caldwell
(1980), has been invoked to account for the morphology-density relation
of Dressler (1980) and the apparent blueing of the galaxy population in
distant clusters (Butcher \& Oemler 1984), and has stimulated
considerable theoretical activity (see Pimbblet 2003 for a review).

Such processes modify the star formation history of the disk by removing
the supplies of gas necessary to fuel further star formation. A
sensitive test of this mechanism would therefore involve a comparison of
the stellar populations of bulges and disks for elliptical, S0 and
spiral galaxies. While the stellar content of galaxies cannot be
resolved except for some of the nearest objects, the radial
distributions of their stellar populations may be examined by means of
colour gradients.

It has long been known that early-type galaxies exhibit negative colour
gradients, in the sense of being bluer at larger radii (Franx,
Illingworth \& Heckman 1989, Peletier et~al.\ 1990, Goudfrooij et~al.\ 
1994, Tamura et~al.\ 1999, Saglia et~al.\ 2000, Tamura \& Ohta 2000,
Hinkley \& Im 2001, Idiart, Michard \& de Freitas Pacheco 2002, Tamura
\& Ohta 2003). These gradients have been conventionally interpreted as
representing gradients in mean metal abundance, as most local
ellipticals also exhibit such gradients in the strength of metal
absorption lines (Baum, Thomsen \& Morgan 1986, Carollo, Danziger \&
Buson 1993, Davies, Sadler \& Peletier 1993). However, this
intepretation is vitiated by the existence of a strong degeneracy
between age and metallicity in broadband colours as well as in most
spectrophotometric indices (Worthey 1996). Nevertheless, colour
gradients have considerable potential for constraining the history of
star formation in distant galaxies, and for providing clues to their
formation and evolution history.

One approach to breaking the age--metallicity degeneracy is to consider
the evolution of colour gradients. This experiment has been attempted in
a few distant clusters with rich populations of early-type galaxies.
Saglia et~al.\ (2000), Tamura \& Ohta (2000) and La Barbera et~al.\ 
(2003) have observed galaxies in clusters at $z \sim 0.4$ and studied
the evolution of their colour gradients, concluding that the most likely
explanation consists of a mild metal abundance gradient within a
passively evolving stellar population formed at very high redshift.

However, because of the relatively limited number of galaxies and
relatively poor resolution at these redshifts, even with Hubble Space
Telescope (HST) imaging, Saglia et al. (2000), Tamura \& Ohta (2000)
and La Barbera et al. (2003) have not examined the dependence of colour 
gradients on galaxy luminosities and morphology and whether the disk and
bulge components of galaxies have different gradients, and therefore 
different stellar populations. It is therefore useful to consider a 
somewhat lower-redshift target, where the galaxy population can be 
probed over a wider range of luminosities and where the higher spatial 
resolution (because of the lower distance) and smaller cosmological dimming 
allow more detailed studies of colour gradient evolution than has been 
possible (cf. Tamura \& Ohta 2003 for a similar approach to the $z=0.03$ 
cluster Abell 2199).

In this paper we discuss a study of colour gradients in the $z=0.18$
cluster Abell 2218 from deep archival HST multicolour data. We describe
the data and the analysis in the following section and present the main
results. We discuss these results in the light of models of galaxy
formation and, particularly, disk evolution. We adopt the concordance
cosmological model with $\Omega_M=0.3$, $\Omega_{\Lambda}= 0.7$ and
H$_0=70$\,km\,s$^{-1}$\,Mpc$^{-1}$.

\section{Data analysis}

\begin{figure*}
\centering\includegraphics[width=150mm]{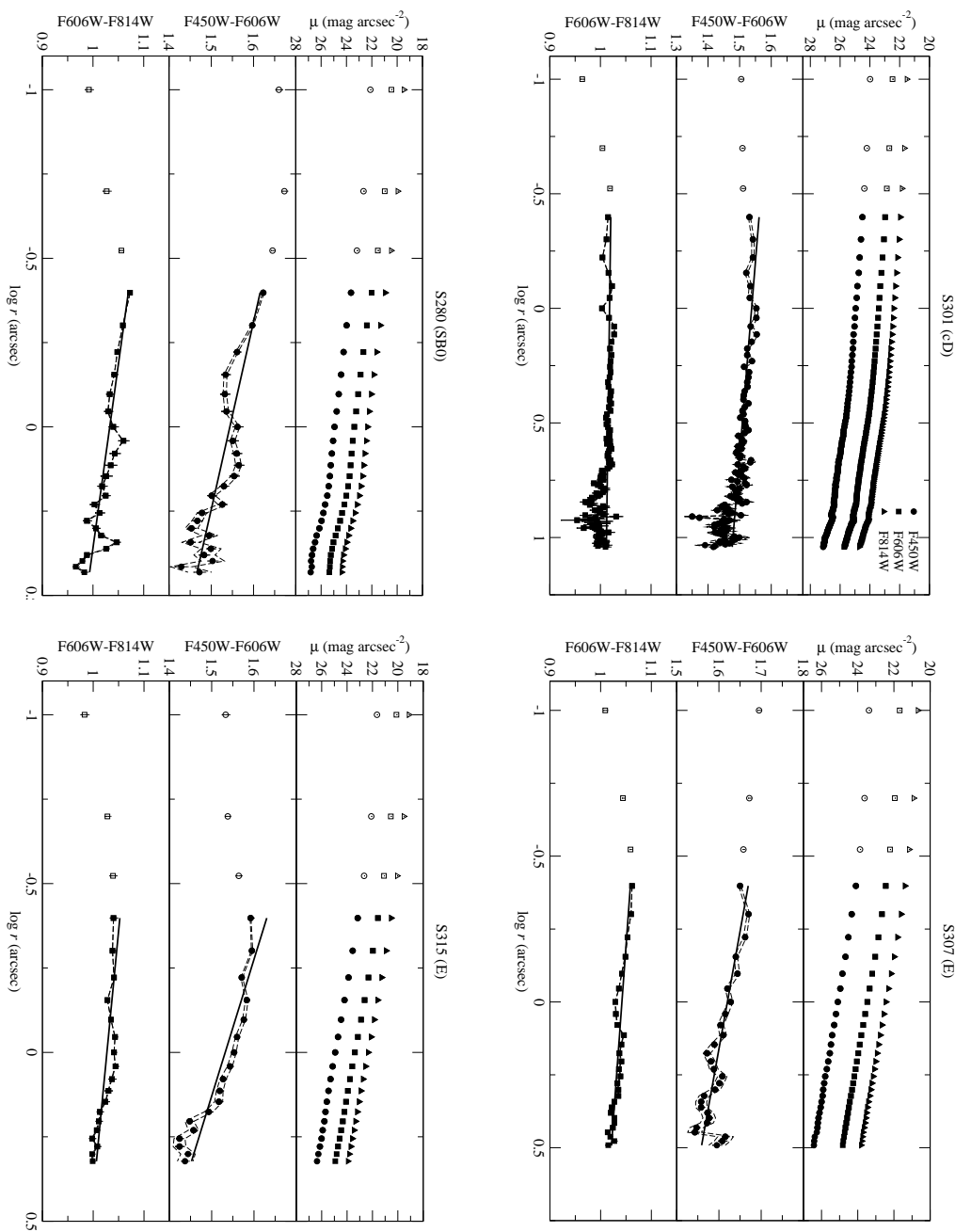}
\caption{Surface brightness profiles and colour gradients for galaxies
  in A2218 (as identified in the figure). In each figure panel (one per
  galaxy), the top panel shows the surface brightness profiles in $B'$
  (circles), $V'$ (squares) and $I'$ (triangles) plotted against $\log r$
  (measured in arcseconds along the semi-major axis); the middle panel shows 
  the colour distribution as a function of $\log r$ in $B'-V'$ (data 
  points are the filled circles, error bars are the statistical error 
  in the flux, the dashed lines show the effect of a $\pm 1 \sigma$ 
  error in subtracting the sky level, and the thick dashed line shows 
  the best fitting straight line); the bottom panel is the same as the 
  middle panel, but for $V'-I'$ and using squares. The open symbols in
  all panels show the surface brightness values and colours for points
  interior to $r = 0.3''$ which were not used in the fit. Note that we
  refer to the $B'$, $V'$ and $I'$ colours by their proper HST names,
  to avoid confusion for those who may wish to make use of these data
  (the actual profiles are available on request in machine-readable
  format).}
\end{figure*}

\addtocounter{figure}{-1}

\begin{figure*}
\centering\includegraphics[width=180mm]{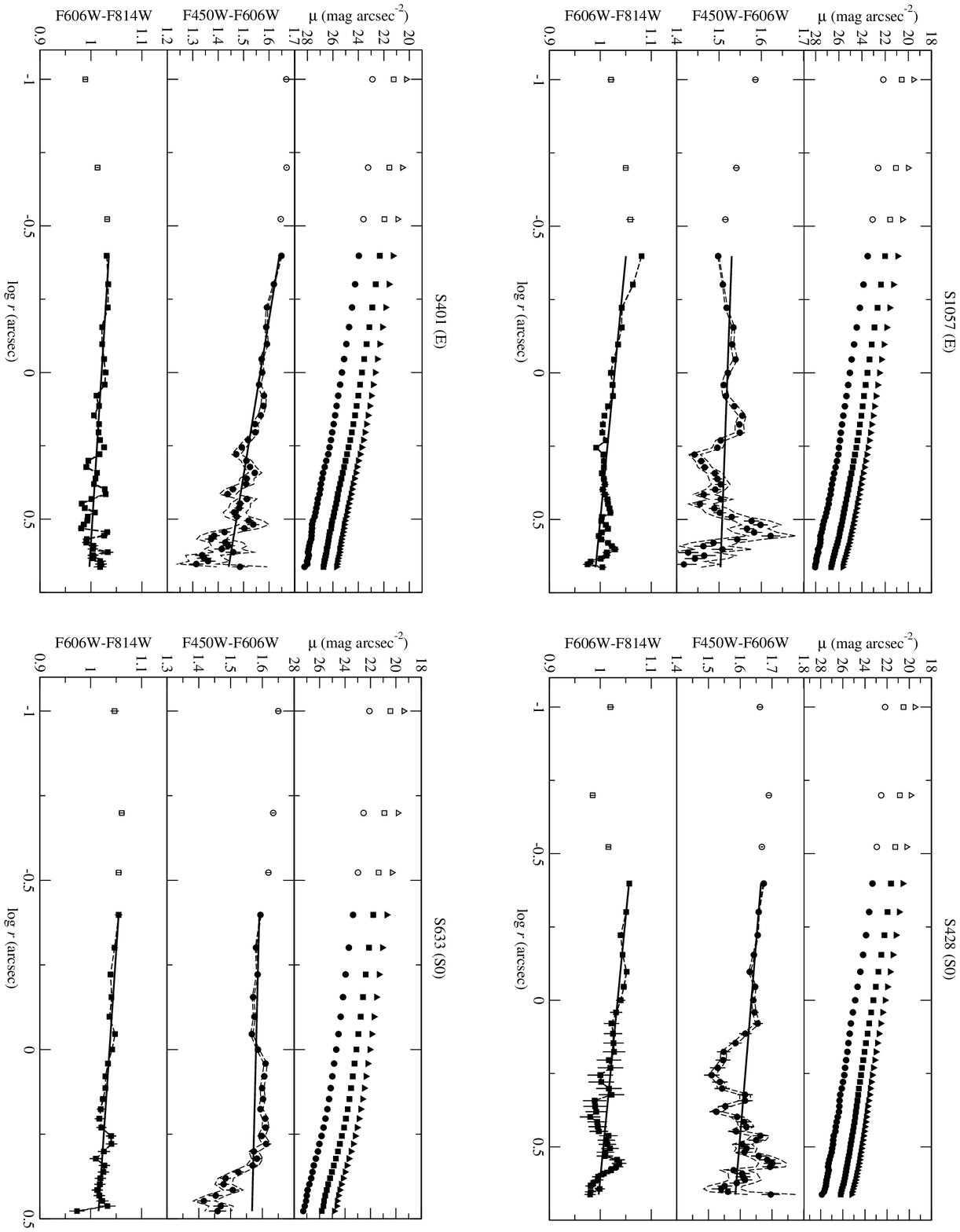}
\caption{(continued)}
\end{figure*}

\addtocounter{figure}{-1}

\begin{figure*}
\centering\includegraphics[width=180mm]{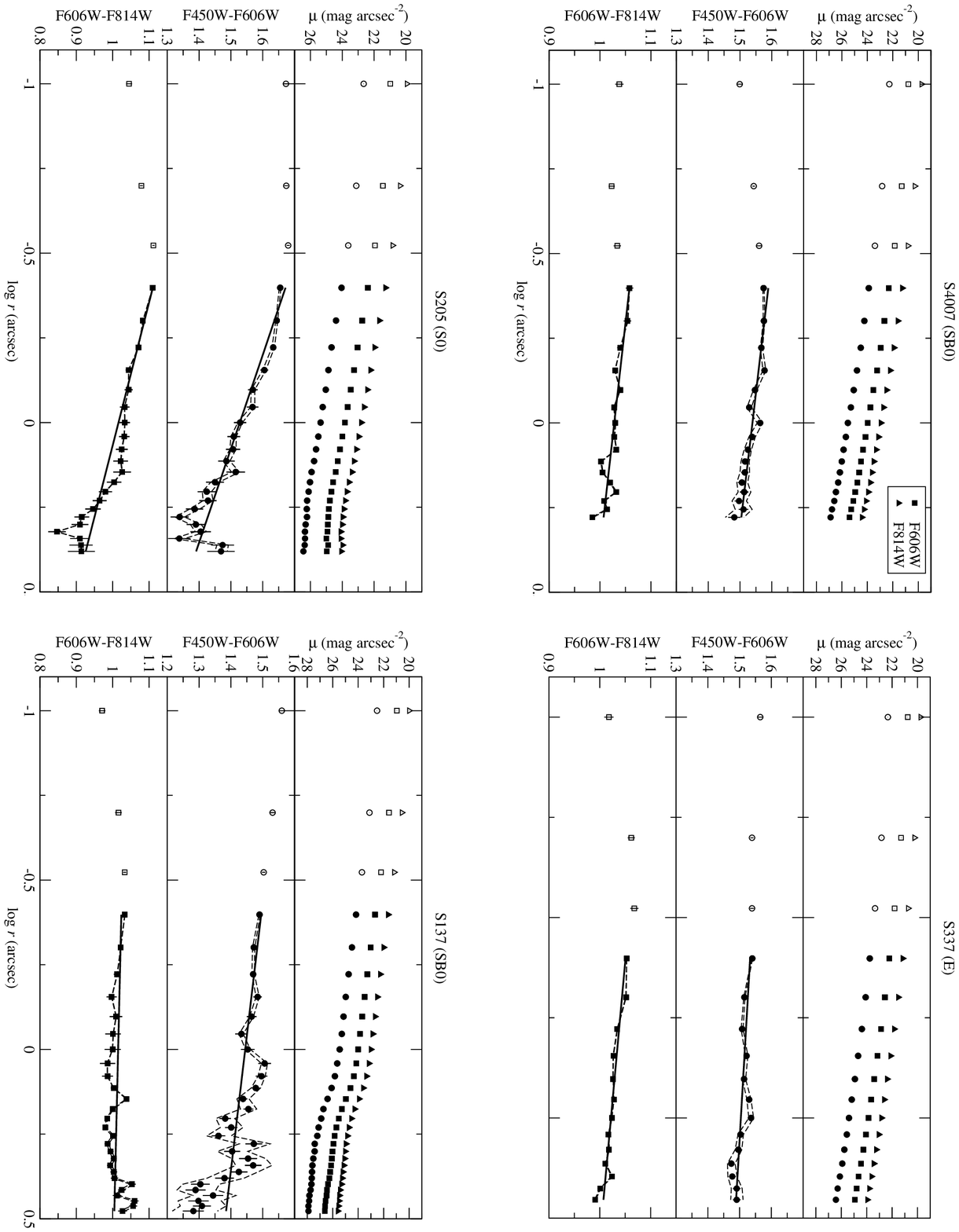}
\caption{(continued)}
\end{figure*}
 
\addtocounter{figure}{-1}

\begin{figure*}
\centering\includegraphics[width=180mm]{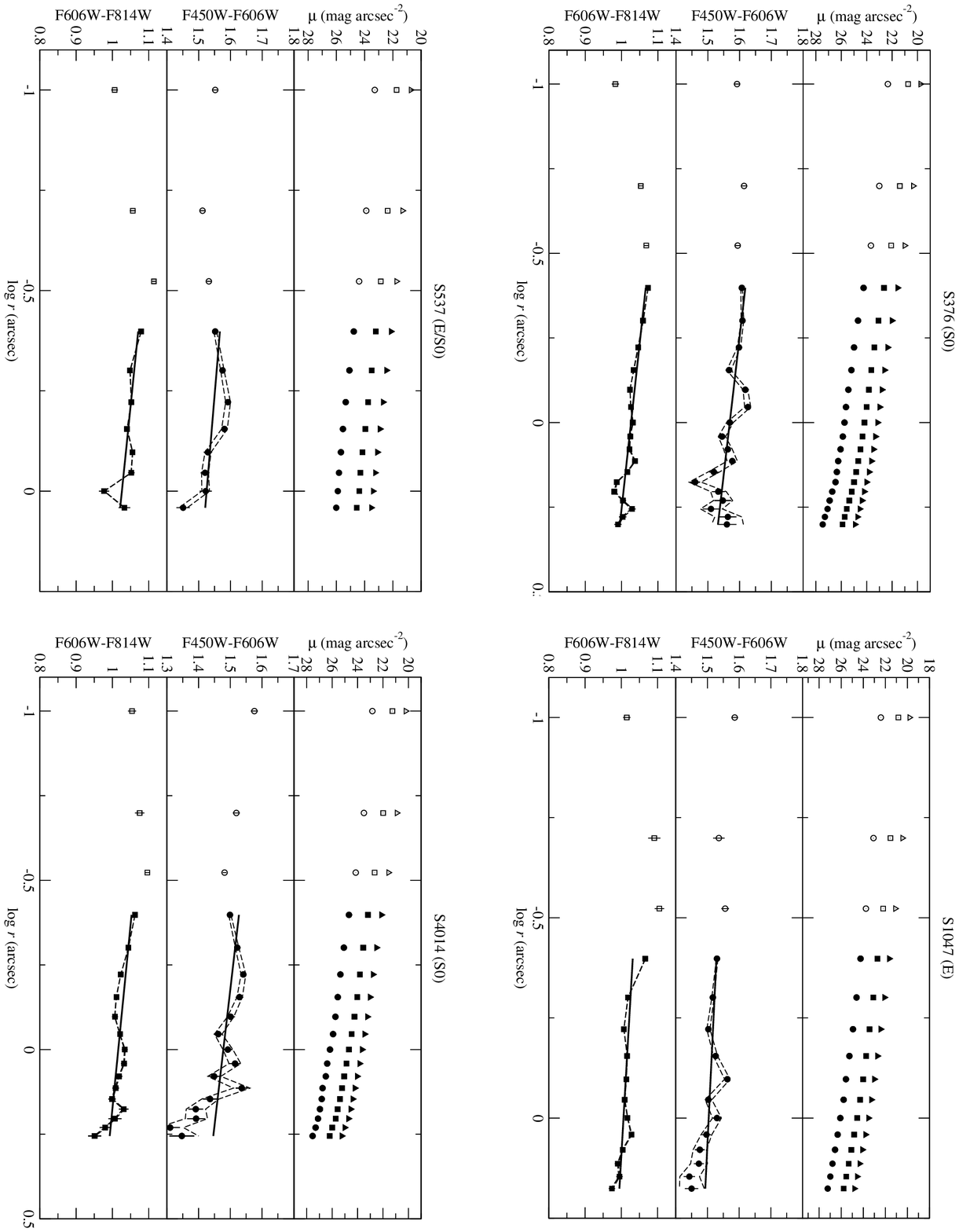}
\caption{(continued)}
\end{figure*}

\addtocounter{figure}{-1}

\begin{figure*}
\centering\includegraphics[width=180mm]{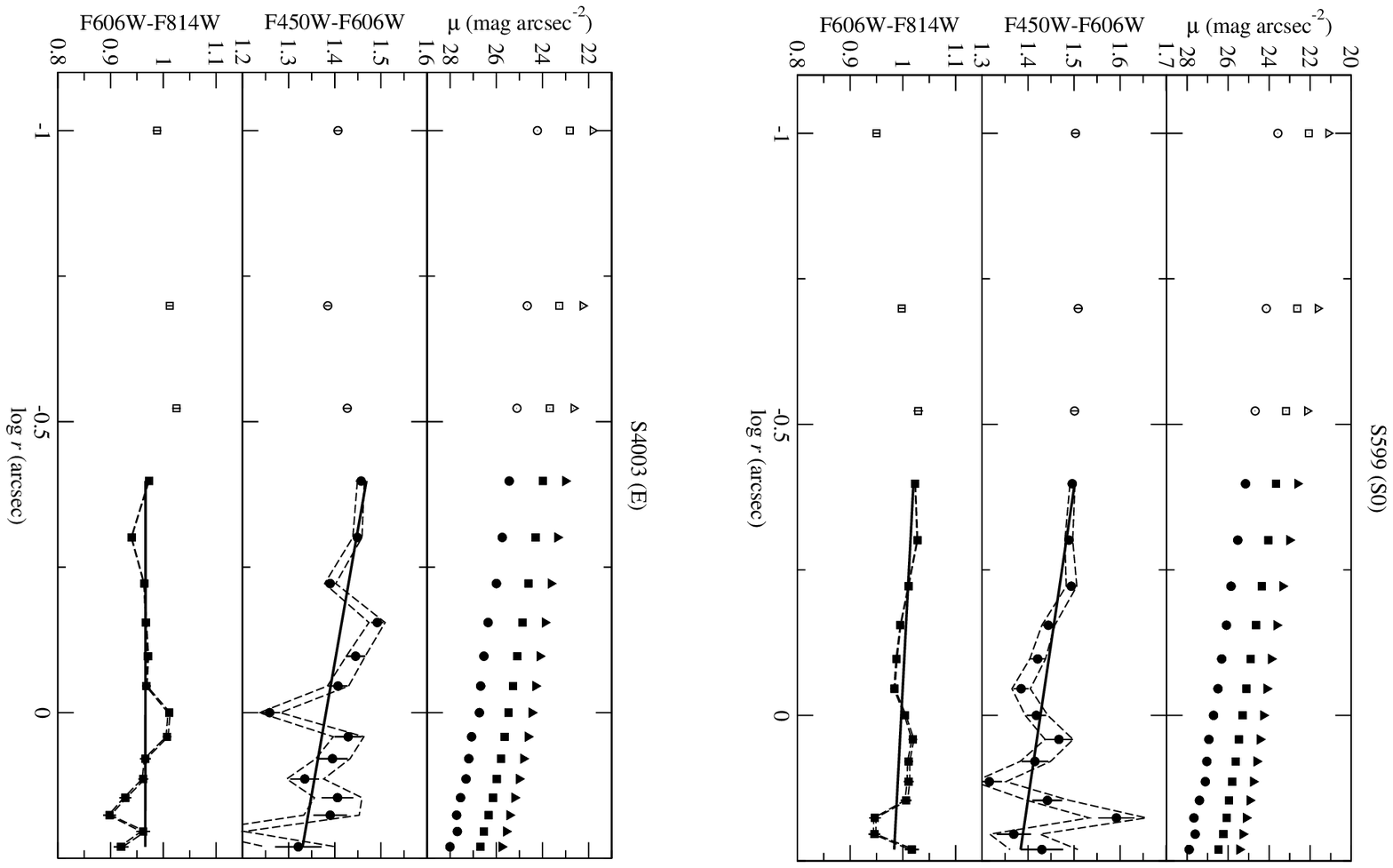}
\caption{(continued)}
\end{figure*}

\vskip 6pt

The data consist of three archival images of the core of Abell 2218
taken by the HST with the Wide Field and Planetary Camera 2 (WFPC2)
through filters F450W, F606W and F814W (PID: 8500; PI: Fruchter). For
simplicity, we will refer to these bands as $B'$, $V'$ and $I'$ in the
remainder of this paper (we use prime symbols to distinguish the bands
from the actual Johnson bands and to avoid confusion in the subsequent
discussion). The total exposure times were 12000s, 10000s and 12000s 
respectively, with individual exposures being 1000s each. These images 
were retrieved as fully processed associations from the HST archive (
Micol, Bristow \& Pirenne 1997). A more detailed description of this 
dataset is given by Smail et~al.\ (2001).

We have selected a sample of 24 galaxies morphologically classified as E
or S0 and spectroscopically confirmed to be cluster members (this
information is taken from Smail et~al.\ 2001). First, the mean sky was
removed from the images using the IRAF task {\tt sky}. For each galaxy
we then derived surface brightness profiles using the IRAF task {\tt 
ellipse} (Jedrzejewski 1987, Busko 1996), using $0.1''$ (1-pixel) sampling 
and 3$\sigma$ clipping to remove overlapping stars and galaxies. As our
main interest in these data is to derive colour gradients, we forced the
isophotes in $V'$ and $I'$ to be evaluated at the same radius (measured
along the semi-major axis) as in the $B'$ image (which is the one with 
the lowest signal to noise) by using the `inellip' option in {\tt ellipse}. 
The isophote centres were left free to be recentered for each image, to 
avoid the effect of minor misalignments between the HST images in the 
different bands.

In some cases, galaxies lie in the envelope of other bright galaxies 
or overlap with nearby galaxies (especially the two giant ellipticals 
but also in a few other cases). In these cases, we have modelled the 
`contaminating' object using the IRAF task {\tt bmodel} and removed it 
from the image. We then computed new isophotes for the galaxies. In most 
cases, this appears to make no difference, at least to the level of 
producing colour gradients different by more than the statistical and
systematic errors. However, galaxies \# 292 and  \#298 appear to be 
significantly affected by the envelopes (and, in the case of \# 292 
proximity to the edge of the detector) of their neighbouring galaxies. 
Because of this we decided to exclude these two galaxies from 
our analysis. In other cases, such as \# 205 and \# 634, modelling shows 
that the profiles are not being affected by proximity to a bright neighbour 
and that {\tt ellipse} has appropriately removed the contaminating object
from the profile..

For each data point, two sources of error contribute; using the terminology
of Saglia et al. (2000) these are: a statistical error, which is the 
uncertainty, returned by {\tt ellipse} in measuring the mean flux along
each isophote and a systematic error, reflecting the uncertainty in 
measuring and subtracting the mean sky level. We estimated this latter
component in the following way: for each of the individual 1000s exposures
in each band, we calculate the mean sky using {\tt sky}. This yields
12, 10 and 12 sky estimates in $B'$,$V'$ and $I'$. The error in the
mean sky is then computed by jackknife resampling of these sky estimates.
 
We calibrate the data on to the Vega system, using the zeropoints 
provided by Holtzman et al. (1995). At each isophote semi-major 
radius we then compute the resulting colour as a function of radius.

We measured the point spread function using non-saturated stars in
our images. There are, unfortunately, only a few such stars but the 
full width at half maximum appears to be a consistent $0.15''$ in all bands, 
suggesting that differences in the point spread function do not affect 
the derivation of colour gradients. Inspection of the surface brightness 
profiles in Figure 1, shows some deviation from the profile expected for 
a de Vaucouleurs profile in the inner two or three points. For this reason, 
we decide to fit to the points with $r > 0.3''$ in order to avoid any 
residual effects due to small differences in the point spread function.

The colour gradients are then derived by fitting the data (colour vs.
logarithm of the radius measured along the semi-major axis) with a
weighted linear least squares regression, where the weights are the
statistical errors of the colours (added in quadrature). The fits are 
carried out to either the radius at which the  quadratic sum of statistical 
and systematic errors exceeds 10\% or to the radius at which {\tt ellipse} 
stopped integrating the surface brightness profile (even if the errors are 
smaller than 10\%). We also estimate the colour gradient only over the range 
defined by our worst colour ($B'-V'$), for consistency. The statistical error 
on the fit slope and intercept is derived from the error in the fit to the 
data, while the systematic error is determined by carrying out a new fit to 
the data points, after adding and subtracting the $\pm 1\sigma$ error in 
the sky level.

Figure~1 shows the surface brightness profiles in all three filters and
the colour gradients in $B' -V'$ and $V'-I'$, plotted against
$\log r$, for all 22 galaxies. The surface brightness profiles and
colour profiles are plotted over the range $r=0$ to the largest radius 
used in fitting the colour gradients. As for points interior to $0.3''$,
which are not used in fitting, some galaxies show colours consistent with
the extrapolated fits, while in some other cases it is possible that the
galaxies contain `cores' with flatter colours. However, this is difficult
to ascertain without data at better resolution.

The major contribution to the error budget comes, not unexpectedly, from 
the $B'$ band, where the WFPC2 camera is less efficient and the galaxies 
are fainter (as it corresponds to the rest-frame $U$). We tabulate the 
results in Table~1. This shows, in column order, the galaxy ID, morphology 
and $K$ band luminosity (from Smail et~al. 2001), the $B' - V'$ colour 
gradient with its statistical and systematic error, and the and $V' - I'$ 
colour gradient with its statistical and systematic error (derived as 
described above). The mean gradients are $\Delta (B' - V') / \Delta \log r 
=-0.14 \pm 0.08$ for Es and $-0.16 \pm 0.09$ for S0s and $\Delta (V'-I') / 
\Delta \log r = -0.06 \pm 0.04$ for Es and $-0.08 \pm 0.06$ for S0s.

\begin{table*}
\caption{Colour gradients}
\begin{tabular}{ccccccccccccccc}
\hline
ID & Morphology & $K$ & $\Delta (B'-V')/d \log r$ & $\sigma_{stat}$ &
$\sigma_{sys}$ & $\Delta (V'-I')/d \log r$ & $\sigma_{stat}$ &
$\sigma{sys}$ \\
\hline
\hline
~301 & cD  & 13.19 & $-$0.059 & 0.002 & 0.016 & $-$0.006 & 0.001 & 0.001 \\
~307 & E   & 13.77 & $-$0.124 & 0.004 & 0.017 & $-$0.042 & 0.002 & 0.001 \\
~280 & SB0a& 14.37 & $-$0.180 & 0.006 & 0.024 & $-$0.092 & 0.004 & 0.002 \\
~315 & E   & 14.55 & $-$0.247 & 0.007 & 0.022 & $-$0.064 & 0.006 & 0.002 \\
1057 & E   & 14.57 & $-$0.026 & 0.006 & 0.042 & $-$0.056 & 0.002 & 0.004 \\
~428 & S0  & 14.60 & $-$0.076 & 0.006 & 0.040 & $-$0.056 & 0.003 & 0.002 \\
~401 & E   & 14.85 & $-$0.188 & 0.005 & 0.044 & $-$0.036 & 0.002 & 0.003 \\
~633 & S0  & 14.88 & $-$0.026 & 0.001 & 0.025 & $-$0.045 & 0.006 & 0.002 \\
4007 & SB0 & 15.17 & $-$0.125 & 0.011 & 0.030 & $-$0.075 & 0.007 & 0.002 \\
~337 & E   & 15.20 & $-$0.073 & 0.009 & 0.016 & $-$0.072 & 0.005 & 0.001 \\
~205 & S0  & 15.21 & $-$0.361 & 0.015 & 0.020 & $-$0.237 & 0.011 & 0.001 \\
~137 & SB0 & 15.52 & $-$0.126 & 0.011 & 0.062 & $-$0.022 & 0.005 & 0.006 \\
~376 & S0  & 15.54 & $-$0.123 & 0.012 & 0.032 & $-$0.102 & 0.006 & 0.002 \\
1047 & E   & 15.77 & $-$0.061 & 0.017 & 0.044 & $-$0.064 & 0.010 & 0.002 \\
~537 & E   & 16.05 & $-$0.105 & 0.023 & 0.020 & $-$0.113 & 0.016 & 0.001 \\
4014 & S0  & 16.22 & $-$0.124 & 0.018 & 0.045 & $-$0.092 & 0.009 & 0.003 \\
~449 & E   & 16.27 & $-$0.235 & 0.010 & 0.031 & $-$0.034 & 0.005 & 0.001 \\
~154 & E   & 16.54 & $-$0.178 & 0.024 & 0.075 & $-$0.072 & 0.011 & 0.006 \\
~612 & S0a & 16.55 & $-$0.230 & 0.018 & 0.056 & $-$0.024 & 0.007 & 0.002 \\
~131 &  E  & 16.93 & $-$0.212 & 0.019 & 0.075 & $-$0.140 & 0.008 & 0.003 \\
~599 & S0  & 17.32 & $-$0.187 & 0.024 & 0.065 & $-$0.058 & 0.011 & 0.004 \\
4003 &  E  & 17.46 & $-$0.220 & 0.028 & 0.071 & $-$0.001 & 0.012 & 0.005 \\
\hline
\end{tabular}
\vskip 1cm
\end{table*}

\section{Results}

In order to model the stellar populations of the galaxies and derive
gradients in age and/or metal abundance from the observed colour gradients,
we need to determine the central colours as a boundary for the models.
We plot these colours in a colour-colour diagram in Figure 2 (where the
colour at $r=0$ is extrapolated from the linear fits), together with the
predictions for a single-age stellar population with present age of 12.5 Gyr 
and metallicity varying between [Fe/H]=$+$0.55 and $-$2.5, and for a 
single-metallicity stellar population with metallicity of [Fe/H]=$+$0.55 
and ages from 12.5 to 2.5~Gyr, as observed through the HST filters and at 
$z=0.18$. We used the latest version of the GALAXEV models of Bruzual \& 
Charlot (2003) to realize these simulations adopting the Padova 1994 
isochrones and Chabrier initial mass function, as recommended by Bruzual 
\& Charlot (2003) and in the GALAXEV documentation.

One possible problem is that, as shown in Figure 2, the models are not
fully capable of reproducing the central colours. This may be due to
some small discrepancy in the population synthesis models, or to some
inaccuracies in the extrapolation of central colours from the $r > 0.3''$
data, or to the presence of dust in the galaxy cores (e.g. Peletier et
al. 1999). However, the reddest (oldest and more metal rich) models are
reasonably close to the actual values of the colours, especially in
the $B'-V'$ colour which is most sensitive to stellar populations and
most useful for our study, and we adopt this as the starting point for
our modelling in interpreting the derived colour gradients.

\begin{figure}
\centering\includegraphics[width=80mm]{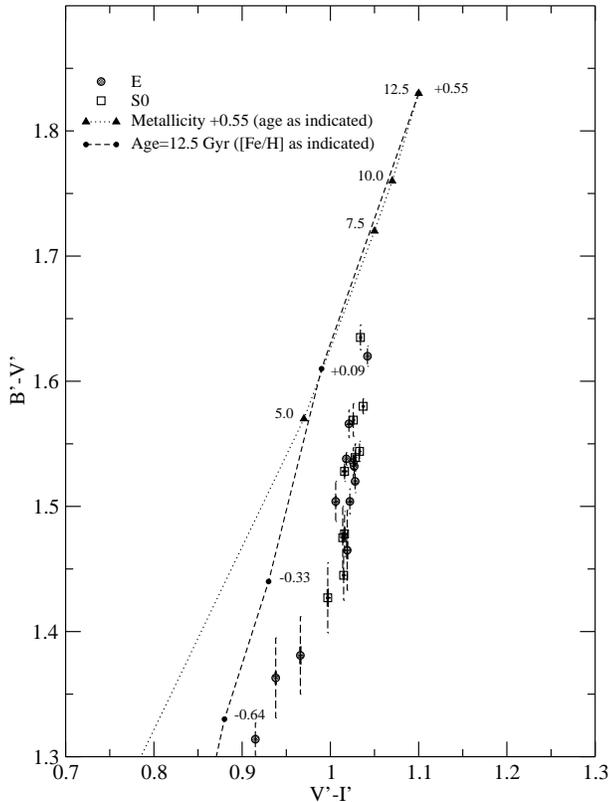}
\caption{Colour--colour plot for E (circles) and S0 (squares) galaxies
  and two BC2003 model lines, as identified in the legend. The values
  presented here are the central colours extrapolated from the fit to
  the data points presented in Figure 1. Solid error bars represent the 
  statistical error while the dashed error bars show the systematic 
  error.}
\end{figure}

One of the aims of this paper is to investigate not only the evolution
of the colour gradients at intermediate redshift and discuss them in
the light of both local and higher redshift data (see below) but also
to consider how the gradient size depends on galaxy luminosity and on
morphology. We plot the derived colour gradients vs. $K$ magnitude (which
is a good measure of the underlying stellar mass) in Figure~3. We use
different symbols (circles and squares, as indicated in the caption)
for E and S0 galaxies. We find little evidence that gradient size, in
both colours, depends on either luminosity or morphology. The consequences
of this finding are examined below.

\begin{figure}
\centering
\includegraphics[width=80mm]{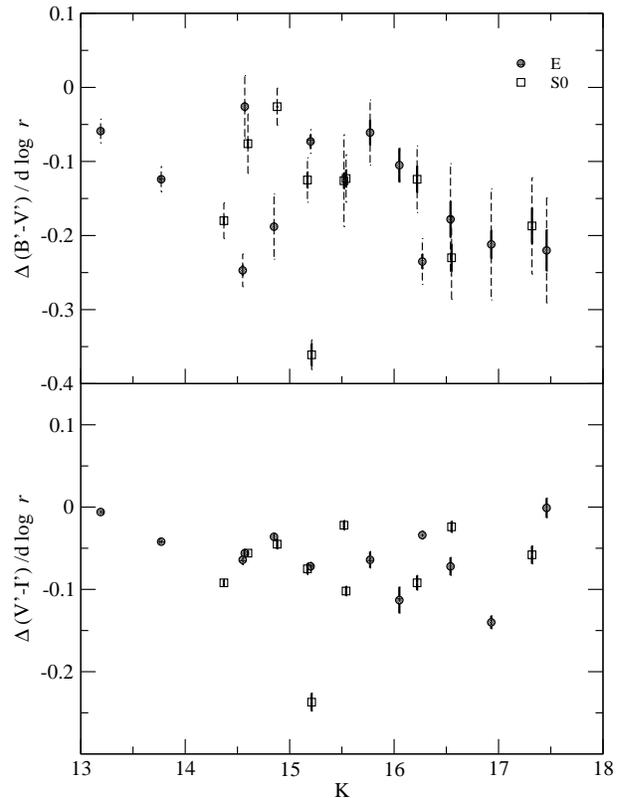}
\caption{Amplitude of the colour gradients in F450W$-$F606W (top panel)
  and F606W$-$F814W (bottom panel) as a function of $K$ magnitude. Symbols
  as in Fig.~2.}
\end{figure}

Naturally, our conclusions prevalently concern spheroids and the bulge
components of S0 galaxies as our data, especially in $B'$ where we are
most sensitive to stellar populations, do not reach much into the disk
dominated portion of the S0 galaxies.  However, in some cases, Figure 1
shows that we are able to sample a small portion of the disk. This may
be the case for \# 280, 633, 205 and 137. It is apparent that the colours
of these disks are at least qualitatively similar to those of their parent
bulges, which would suggest that they have similar stellar populations.
Fisher, Franx \& Illingworth (1996) and Peletier \& Balcells (1996) have
also reached similar conclusions, although Fisher et al. (1996) also point
out that the star formation histories may be different for at least some
galaxies, despite their having similar ages. 

Unfortunately, the small portions of the disks we survey are not sufficient
to determine colour gradients for the disks and carry out a comparison with
the bulge and therefore reconstruct the star formation and enrichment histories
of the two components of the galaxies. This should be possible with more
sensitive data from the ACS, some of which are available publicly and will
be the subject of a later paper.

\section{Discussion}

We now examine the implications of our findings for models of galaxy
formation and evolution: in particular, we consider simplistic models
of formation by dissipational collapse and by hierarchical mergers
as our benchmarks; these comparisons should probably be intended as
a stimulus for more comprehensive modelling rather than setting true
limits on the accuracy with which the models represent the observations

We first find that the size of the colour gradients is broadly independent 
of $K$ luminosity. Models of colour gradients in galaxies, unfortunately, 
do not yet reach this level of detail, but we naively expect that, in a 
monolithic collapse scenario, the size of the gradients will be proportional
to the depth of the potential well, as the stellar population gradients are
induced by superwinds, so that more massive galaxies have steeper colour
gradients (Carlberg 1984). In general, monolithic collapse models yield 
excessively large gradients when compared to the observations (Carlberg 1984) 
but more detailed treatment of gas physics may alleviate this problem (e.g.
Kawata 2001, Pipino \& Matteucci 2004). In hierarchical merger models
(White 1980, Kauffmann 1996, Cole et~al.\ 2000) we expect that mergers will 
erase any colour gradients originally set up. Some gradients may be 
re-established by star formation at later epochs, but we might expect that 
more massive galaxies, at the top of the merging hierarchy, will experience 
more mergers and have flatter gradients. Our results appear to be inconsistent
with either (admittedly simplistic) scenario, as we observe no evidence of 
strong trends in gradient size with luminosity. 

Tamura \& Ohta (2003) suggest that the brigher ($R < 15$) and more massive 
galaxies in Abell 2199 show steeper gradients, which would be consistent 
with the predictions of monolithic collapse models, while La Barbera et al.
(2004) favour no trend in the gradient size with luminosity, or possibly 
flatter gradients for brighter galaxies, in Abell 2163B, at $z \sim 0.2$ 
which is in somewhat better agreement with our observations and with 
hierarchical merger models. Study of the evolution of these trends
with redshift may allow us to resolve this issue more decisively.

The observations suggest that there is no significant difference in the
size of colour gradients between E and S0 galaxies bulges. This implies 
that the two systems have very similar distributions of stellar populations. 
We use our single stellar population models to calculate that the observed
gradients are consistent with either a $\delta{\rm[Fe/H]}/\delta\log r
\sim -0.3$ dex per decade gradient at a fixed 12.5~Gyr (present) age, or
with an approximately 3~Gyr per decade age gradient at a fixed
metallicity of [Fe/H]=+0.09, consistent with most previous work on this
subject. The age--metallicity degeneracy, of course, prevents us from
determining, from these data alone, which scenario is most likely.

We therefore follow previous work (Saglia et~al.\ 2000, Tamura \& Ohta
2000, La Barbera et~al.\ 2003) in studying the evolution of colour
gradients as a function of lookback time. Our $B'$, $V'$ and $I'$
bands correspond approximately to the rest-frame $U$, $g$ and $R$ bands.
Since few local studies are carried out in the $g$ band, we compare our
results with observations in $U-V$ and $V-R$. Local values for $U-V$ and
$V-R$ gradients for cluster galaxies are taken from the compilation of
Idiart et~al.\ (2002). We use the $U-V$ gradients for a $z=0.38$ cluster
presented in Saglia et~al.\ (2000); we have been unable to find $V-R$
data at higher redshift.

We plot these data and the mean values for colour gradients in A2218 in
Figure~4, together with four models from La Barbera et~al.\ (2003) that
appear to be most appropriate to reproduce their observations. The
parameters of these models are shown in Table~2, where the first column
is the model ID and the subsequent columns represent the age at centre
and outskirts, the metal abundance at centre and outskirts and the
exponential decay time of the star formation. Here, model T1 has a
central age of 13 Gyr and an outer age of 8 Gyr, with both stellar
populations having a metallicity of $+0.09$: it corresponds to the
model of the same name in La Barbera et al. (2003).  Model Z1 has
two stellar popularions of the same age (13 Gyr) but a metallicity
gradient of 0.4 dex from center to edge. Model TZ has both a small
age gradient (13 Gyr at the center and 11 at the edge) and a metal
abundance gradient as in model T1. Finally, model T3 corresponds to
La Barbera's model $\sim T1$ (for spirals), has a small age gradient 
(2 Gyr from center to edge, with a central age of 13 Gyr) and metal 
abundance of $-0.33$ but $\tau$ of 3 Gyr, while all other models have 
$\tau$ of 1 Gyr (where $\tau$ is the exponential decay time for star 
formation). The actual values for the ages and metallicities of the models 
are slightly different from those used in La Barbera et al. (2003), as
the cosmology and metallicity have changed in the latest GALAXEV models. 
We use the GALAXEV values that best approximate those used by La Barbera 
et al. (2003).

The models are unable to reproduce the central colours (Figure 2),
which suggests that these galaxies may be more metal rich than 
[Fe/H]=$+0.55$ (this is likely due to the typical $\alpha$ element
overabundances). Because of this, there is some doubt as to whether
the derived gradients may be fairly compared to the observations.
The actual values of the colour gradients from the models at $z=0.18$
are, for models T1,Z1,TZ and T3, respectively, $- 0.40$, $-0.28$, $-0.35$
and $-0.19$ mag. in $U-V$ and $-0.08$, $-0.06$, $-0.08$ and $-0.05$ mag. 
in $V-R$. While all models are able to reproduce the $V-R$ gradients, 
they are a poor fit to the $U-V$ gradient, with the possible exception 
of model T3. However, if we consider the evolution of the gradients, model T3 
predicts a gradient of $-0.13$ at $z=0$ and $-0.30$ at $z=0.38$, which
compares badly with the measured values of $-0.17$ at $z=0$ (Idiart et
al. 2002) and $-0.08$ at $z=0.38$ (Saglia et al. 2000). Since the models
do not appear to reproduce the values of the gradients or the galaxy
central colours, we choose to carry out a differential comparison by
forcing the models in Figure 4 to run through the A2218 measurements. 

\begin{figure}
\centering\includegraphics[width=90mm]{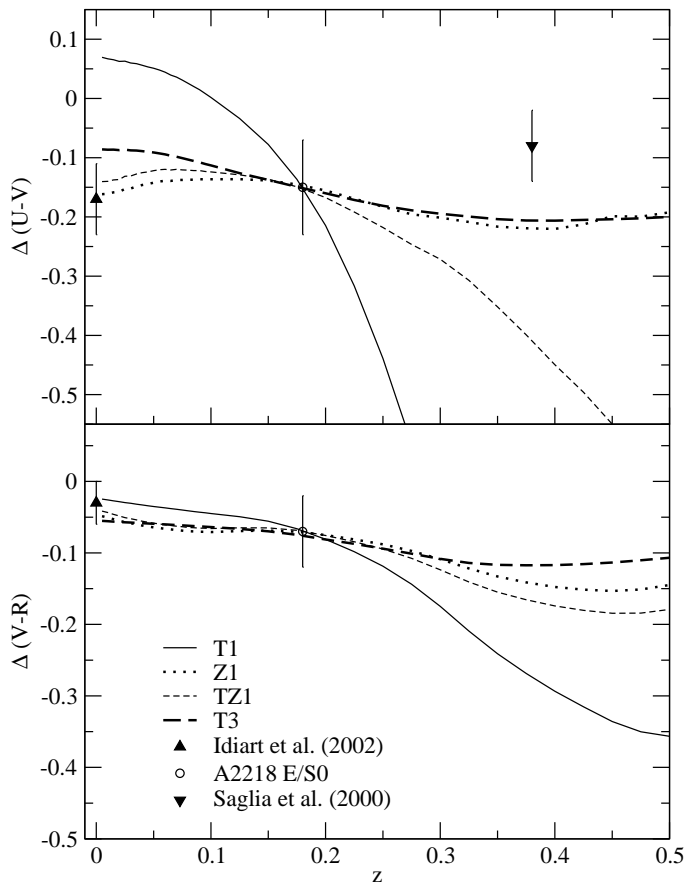}
\caption{Mean colour gradients in $B'-V'$ (top panel) and
  $V'-I'$ (bottom panel) for A2218 galaxies (circles for Es and
  squares for S0s; note the arbitrary offset from the cluster redshift
  introduced for purposes of clarity in presentation). Also plotted are
  literature values (Filled `up' triangles: Idiart et al. 2002; filled 
  `down' triangles: Saglia et al. 2000) and four models whose parameters 
  are described in Table~2 and in the text.}
\end{figure}

The main conclusion from this exercise is that the best model is the
familiar small metallicity gradient (although a slightly smaller
gradient may perform better), while a small age gradient model with
large $\tau$ is marginally worse. Note that this latter model performs
somewhat better in reproducing the actual values of the gradients. The 
combined age and metallicity model and the age gradient model are largely 
excluded by the data.

The similarity in the size of colour gradients for E and S0 galaxies
suggests that their bulges, at least, have similar stellar populations 
and that these are also similarly distributed. The result confirms and 
extends the observations of Ziegler et~al.\ (2001) and Mehlert et~al.\ 
(2003) for bright S0 galaxies in Abell 2218 and the  Coma cluster 
respectively: on the other hand, both Kunstchner \& Davies (1998) and 
Mehlert et al. (2003) find that at least some S0s in Fornax and Coma, 
respectively, possess younger stellar populations, although these objects 
tend to be among the fainter S0s rather than the bright S0s. Similarly 
Smail et al. (2001) argue that the fainter S0s in their sample ($K > 17$,
or $L < 0.1 L^*$) may be more distinct from ellipticals. This need not be 
inconsistent with our findings, if bright S0s form in a different manner 
from the faint S0s: in particular, the evolution may take place mostly in 
the disks, while our results concern mostly bulges. 

Our main conclusions are therefore that the observed trends in colour 
gradients are difficult to explain in the context of any of the popular 
models of galaxy formation, but stress the essential similarity of the 
stellar populations of bulge dominated systems. Observations of a larger 
sample of objects at higher resolution with ACS or WFPC3 should allow us 
to investigate these issues further. Investigations of colour gradients 
for S0 disks should allow us to achieve a better understanding of the 
relation between bulge and disk evolution, particularly with reference 
to the scenario in which spiral disks are converted to S0 disks via a 
variety of processes (ram stripping, truncation, starbursts induced by 
mergers or interactions, harassment, among others). 

\begin{table*}
\caption{Model parameters}
\begin{tabular}{cccccc}
Model & Age at centre & Age at outskirts & [Fe/H] at centre &
[Fe/H] at outskirts & $\tau$ \\
\hline
\hline
T1 & 13 & ~8 & $+$0.09 & $+$0.09 & 1 \\
Z1 & 13 & 13 & $+$0.09 & $-$0.33 & 1 \\
TZ & 13 & 11 & $+$0.09 & $-$0.33 & 1 \\
T3 & 13 & 11 & $-$0.33 & $-$0.33 & 3 \\
\hline
\end{tabular}
\vskip 1cm
\end{table*}

\section*{Acknowledgements}

We wish to thank the anonymous referee for a very helpful and comprehensive
report, which has substantially helped in improving this paper and in
clarifying our presentation.

\setlength{\bibhang}{2.0em}

\clearpage
\label{lastpage}


\begin{thebibliography}{50}

\setlength{\itemindent}{-2.5em}

\expandafter\ifx\csname natexlab\endcsname\relax\def\natexlab#1{#1}\fi

\bibitem[Abraham et al.(1999)]{ab99} Abraham, R. G., Ellis, R. S.,
  Fabian, A. C., Tanvir N. R., Glazebrook, K. 1999, MNRAS, 303, 641
\bibitem[Baum et al.(1986)]{ba86} Baum, W. A., Thomsen, B. and Morgan,
  B. L. 1986, ApJ, 301, 83
\bibitem[Bruzual and Charlot(2003)]{bc03} Bruzual A., G. and Charlot, S.
  2003, MNRAS, 344, 1000
\bibitem[Busko(1996)]{bu96} Busko, I. C. 1996, Astronomical Data
  Analysis Software and System V, ASP Conference Series 101, G. H.
  Jacoby and J. Barnes, p. 139
\bibitem[Butcher and Oemler(1984)]{bo84} Butcher, H. and Oemler, A.
  1984, ApJ, 285, 426
\bibitem[Carlberg(1984)]{ca84} Carlberg, R. 1984, ApJ, 286, 403
\bibitem[Carollo et al.(1993)]{ca93} Carollo C. M., Danziger, I. J.  and
  Buson, L. 1993, MNRAS, 265, 553
\bibitem[Cole et al.(2000)]{col00} Cole S., Lacey C. G., Baugh C. M.,
  Frenk C. S. 2000, MNRAS, 319, 168
\bibitem[Davies et al.(1993)]{da93} Davies, R. L., Sadler, E. M. and
  Peletier, R. F. 1993, MNRAS, 262, 650
\bibitem[Dressler(1980)]{dr80} Dressler, A. 1980, ApJ, 236, 351
\bibitem[Fisher et al. (1996)]{fi96} Fisher D., Franx M. and Illingworth,
G. D. 1995, ApJ, 459, 110
\bibitem[Franx et al.(1989)]{fr89} Franx, M., Illingworth, G.  and
  Heckman, T. 1989, AJ, 98, 538
\bibitem[Goudfrooij et al.(1994)]{go94} Goudfrooij, P., Hansen, L.,
  J{\o}rgensen, H. E., N{\o}rgaard-Nielsen, H. U., de Jong, T. and van
  den Hook, L. B. 1994, A\&AS, 104, 179
\bibitem[Hubble(1936)]{hub36} Hubble E. P. 1936, {\it The Realm of the
  Nebulae}, Yale University Press
\bibitem[Hinkley and Im(2001)]{hi01} Hinkley, S. and Im, M. 2001, ApJ,
  560, L41
\bibitem[Holtzman et al.(1995)]{ho95} Holtzman, J. A., Burrows, C. J.,
  Casertano, S., Hester, J., Trauger, J. T., Watson, A. M. and Worthey,
  G. 1995, PASP, 107, 1065
\bibitem[Idiart et al.(2002)]{id02} Idiart, T. P., Michard, R. and de
  Freitas Pacheco, J. A. 2002, A\&A, 383, 30
\bibitem[Jedrzejewski(1987)]{je87} Jedrzejewski, R. L. 1987, MNRAS, 226,
  747
\bibitem[Kauffmann(1996)]{kau96} Kauffmann G. 1996, MNRAS, 281, 487
\bibitem[Kawata(2001)]{kaw01} Kawata, D. 2001, ApJ, 558, 598
\bibitem[Kormendy and Bender(1996)]{kb96} Kormendy J. and Bender R. 1996,
ApJ, 464, L119
\bibitem[Kunstchner and Davies(1998)]{kd98} Kunstchner H. and Davies R. L.
1998, MNRAS, 295, 43
\bibitem[La Barbera et al.(2003)]{la03} La Barbera, F., Busarello, G.,
  Massarotti, M., Merluzzi, P. and Mercurio, A. 2003, A\&A, 409, 21
\bibitem[La Barbera et al.(2004)]{la04} La Barbera, F., Merluzzi P.,
  Busarello, G., Massarotti, M. and Mercurio, A. 2004, A\&A, 425, 797
\bibitem[Larson et al.(1980)]{la80} Larson, R. B., Tinsley, B. M. and
  Caldwell, C. N. 1980, ApJ, 237, 692
\bibitem[Mehlert et al.(2003)]{me03} Mehlert, D., Thomas, D., Saglia, R.
  P., Bender, R., Wegner, G. 2003, A\&A, 407, 423
\bibitem[Micol et al.(1997)]{mic97} Micol A., Bristow P., Pirenne B.
  1997, in {\it The 1997 HST Calibration Workshop with a new generation
  of instruments} ed. S. Casertano, R. Jedrzejewski, C. D. Keyes and
  M. Stevens (Baltimore: Space Telescope Science Institute)
\bibitem[Michard(1994)]{mi94} Michard R. 1994, A\&A, 288, 401
\bibitem[Peletier et al.(1990)]{pe90} Peletier, R. F., Davies, R. L.,
  Illingworth, G. D., Davis, L. E. and Cawson, M. 1990, AJ, 100, 1091
\bibitem[Peletier and Balcells(1996)]{pb96} Peletier R. and Balcells
M. 1996, AJ, 111, 2238
\bibitem[Peletier et al.(1999)]{pe99} Peletier, R. F., Balcells, M.,
Davies, R. L., Andredakis Y., Vazdekis, A., Burkert, A. and Prada, F.
1999, MNRAS, 310, 703
\bibitem[Pimbblet(2003)]{pim03} Pimbblet K. A. 2003, PASA, 20, 294 
\bibitem[Pipino and Matteucci(2004)]{pm04} Pipino, A. and Matteucci, F.
  2004, MNRAS, 347, 968
\bibitem[Saglia et al.(2000)]{sa00} Saglia, R., Maraston, C., Greggio,
  L., Bender, R. and Ziegler, B. 2000, A\&A, 360, 911
\bibitem[Smail et al.(2001)]{sm01} Smail, I., Kuntschner, H., Kodama,
  T., Smith, G. P., Packam, C., Fruchter, A. S., Hook, R. N. 2001,
  MNRAS, 323, 839
\bibitem[Tamura et al.(2000)]{ta00} Tamura, N., Kobayashi, C., Arimoto,
  N., Kodama, T. and Ohta, K. 2000, AJ, 119, 2134
\bibitem[Tamura and Ohta(2002)]{to00} Tamura, N. and Ohta, K. 2000, AJ,
  120, 533
\bibitem[Tamura and Ohta(2003)]{to03} Tamura, N. and Ohta, K. 2003, AJ,
  126, 596
\bibitem[White(1980)]{wh80} White, S. D. M. 1980, MNRAS, 191, 1
\bibitem[Worthey(1996)]{wo96} Worthey, G. 1996, in New light on galaxy
  evolution, IAU Symposium 171, ed. R. Bender and R. L. Davies
  (Dordrecht: Kluwer), p. 71
\bibitem[Ziegler et al.(2001)]{zi01} Ziegler, B., Bower, R. G., Smail,
  I., Davies, R. L., Lee, D. 2001, MNRAS, 325, 1571

\end{thebibliography}
\end{document}